\documentclass[conference]{IEEEtran}

\IEEEoverridecommandlockouts 

\usepackage{cite}
\usepackage{amsmath,amssymb,amsfonts}
\usepackage{array}
\usepackage{graphicx}
\usepackage{textcomp}
\usepackage{hyperref}
\usepackage{xcolor}
\usepackage{algorithmic}
\usepackage{algorithm}
\usepackage{multirow}
\usepackage{caption}
\usepackage{subcaption}

\begin{document}

\title{\LARGE \bf
An Initial Exploration: Learning to Generate Realistic Audio for Silent Video
}

\author{Matthew Martel, Jackson Wagner
\thanks{mdmartel@berkeley.edu, jackson.wagner@berkeley.edu}%
}

\maketitle
\thispagestyle{empty}
\pagestyle{empty}

\begin{abstract}

Generating realistic audio effects for movies and other media is a challenging task that is accomplished today primarily through physical techniques known as Foley art. Foley artists create sounds with common objects (e.g., boxing gloves, broken glass) in time with video as it is playing to generate captivating audio tracks. In this work, we aim to develop a deep-learning based framework that does much the same - observes video in it's natural sequence and generates realistic audio to accompany it. Notably, we have reason to believe this is achievable due to advancements in realistic audio generation techniques conditioned on other inputs (e.g., Wavenet conditioned on text). We explore several different model architectures to accomplish this task that process both previously-generated audio and video context. These include deep-fusion CNN, dilated Wavenet CNN with visual context, and transformer-based architectures. We find that the transformer-based architecture yields the most promising results, matching low-frequencies to visual patterns effectively, but failing to generate more nuanced waveforms. 

\end{abstract}

\section{INTRODUCTION}

Generating realistic waveforms from visual context is useful for many creative media tasks, like developing soundtracks for animated films or enhancing audio effects for live-action movies. However, the task is not easy for a number of reasons. First, the audio-generation task, by which a system generates audio that mimics that found in a dataset (e.g., speech generation systems), is challenging in itself. These systems have become quite advanced, but are normally geared towards creating a specific type of sound, like human speech, for example \cite{wavenet} \cite{transformer_sound}. To generate realistic sound for video, a wide range of waveforms will need to be generated to mimic those associated with a variety of sources (e.g., car engine vs baby crying). Second, video context is inherently limited in its ability to inform audio generation. For one, a typical video may have only 30 to 60 frames per second, while the associated audio samples may be played back at between 8kHz in the lowest quality cases and 44.1kHz in more typical cases, making it so that large sequences of audio samples must be inferred from unchanging video context. Further, sounds that are not directly related to content displayed in video frames are common (e.g., a siren in the distance, a person talking off-camera). Any system we build will likely be limited in its capacity to predict and generate these sounds. Third, creating video context embeddings that accurately encode visual information that is relevant specifically for the audio generation task poses another challenge. What visual queues lead to audio of what type? How should we handle multiple relevant visual queues present in a single instance of video context? These are questions our system will need to learn to address if it is to perform well. 

We explore three different model architectures to accomplish this task of realistic audio generation from silent video. In all three cases, our key idea is to augment audio-generation models with visual context in the form of embeddings, which is joined with audio generation streams at various stages of the computation. We train our models to output two-channel audio that mimics that found in YouTube videos and home videos recorded by us, given respective visual contexts. 

The first architecture we test is a deep-fusion CNN that processes previously generated audio and video in parallel to output the audio segment associated with the next video frame. Second, we extend the dilated Wavenet CNN architecture \cite{wavenet} by adding a video context embedding to audio context as an initial step in the forward pass. This model outputs the next audio sample rather as opposed to the next audio sequence. Finally, we develop an audio and video transformer architecture that similarly takes in audio context and a video context embedding and generates the next audio sample via a multi-head attention transformer module. We test these architectures on different types of video data using different loss functions and record results.

Our work establishes the audio and video transformer architecture with cross entropy loss a promising approach to audio generation from silent video. It also illustrates the short comings of the other methods we developed, namely the strategy of generating audio for the next frame, and using a dilated Wavenet architecture without extensive training. 

\section{RELATED WORK}
\subsubsection{Audio Generation} 
Audio generation techniques have been extensively developed in recent years. In 2016, the paper "WaveNet: A Generative Model for Raw Audio," by van den Oord et al., introduced a  a deep neural network architecture for generating raw audio waveforms that produced state-of-the-art results when applied to the text-to-speech task. The model leverages a dilated convolutional neural network architecture to generate audio samples sequentially \cite{wavenet}. More recently, in the 2021 paper, "A Generative Model for Raw Audio Using Transformer Architectures," Verma and Chafe demonstrate that a transformer architecture can be effective in generating audio at the waveform level \cite{transformer_sound}. In our work, we leverage both of these findings to inform architectures we develop for waveform generation from silent video. 

\subsubsection{Video Interpretation}
Our work also takes inspiration from developments in video interpretation methods. In the 2021 paper, "CLIP4Caption: CLIP for Video Caption," Tang et al. introduce an approach for video captioning that leverages a CLIP-enhanced video-text matching network. This approach was shown to be effective in encouraging the model to learn video features that are highly correlated with text for text generation \cite{clip_4_vid_caption}. Similarly, models we develop will need to learn video features that are highly correlated with audio for the audio generation task. In the 2022 paper, "End-to-end Dense Video Captioning as Sequence Generation," Zhu et al. propose an approach to video captioning that frames the problem as a sequence generation task using a multi-modal transformer \cite{vid_caption_2}.  We take inspiration from this approach to develop our audio and video transformer architecture, that similarly treats the problem of waveform generation as sequence generation, and leverages a transformer to do so. 

\subsubsection{Image Recognition} The image recognition task is core to the derivation of meaning from video frames, and is therefore also relevant to our work. The 2015 paper, "Deep Residual Learning for Image Recognition," by He et al., is of notable use to us, because it introduces an architecture that mitigates the vanishing-gradient problem, for CNNs in this case, by exploiting residual or identity connections in the forward pass \cite{resnet}. We leverage the findings from this paper and extend the approach to three dimensions (x, y, time) to develop video embeddings for visual context. 

\subsubsection{Previous audio from video work}
Our work is not the first to develop an audio-from-video generation technique. In the 2018 paper, "Visual to Sound: Generating Natural Sound for Videos in the Wild," Zhou et al. propose a technique for generating realistic audio for videos from the visual context. The approach treats the audio generation task as a conditional generation problem where the goal is to estimate $p(y1, y2, ..., yn | x1, x2, ..., xn)$ where $yi$ are audio samples and $xi$ are video frames. Models this paper develops consist of two parts: video encoders and a sound generator. For their sound generator, the researchers use a SampleRNN architecture \cite{sample_rnn}. For video encoders, they experiment with a frame-to-frame based based method, sequence to sequence method, and finally and optical flow based method to capture movement \cite{visual_to_sound}. In our work, we take a sequence-to-sequence approach for video encoding, presenting our model with an embedding that captures the last $n$ video frames when generating the next audio sample. We extend the work presented here by introducing different audio-generation techniques, namely in the forms of deep-fusion, dilated Wavenet CNN, and transformer architectures. 


\section{METHODS}
Our method for audio generation from silent video leverages a deep-learning approach. At test-time, our models are presented with both audio context (previously generated audio or initial audio sample) and video context (the past $n$ frames of the video), and are tasked with generating audio associated with the most recently observed video frame in a sequential manner. Find an illustration of this process in Figure \ref{fig:generation}. 

\subsection{Data sources}
Our work utilizes two primary sources of data, videos from YouTube and homemade videos collected by the authors. Videos from YouTube are downloaded using the Python pytube library and homemade videos are collected using a smartphone.\\

\begin{raggedright}\begin{tabular}{ |p{1.75cm}||p{1.75cm}|p{1.5cm}|p{1.35cm}| p{1cm}| }
 \hline
 \multicolumn{5}{|c|}{Data Parameters} \\
 \hline
 Data Source & Video \newline Resolution & Audio Frequency & Audio Channels & Format\\
 \hline \hline
 YouTube   &  640x360x3   & 44.1 kHz &   2 & MP4\\
 \hline
 Homemade &   1920x1080x3  & 44.1 kHz   & 2 & MOV\\
 \hline
\end{tabular}\end{raggedright}\\

\begin{figure}[h!]
    \centering
    \begin{subfigure}[b]{.2\textwidth}
    \centering
            \fbox{\includegraphics[height = 2cm]{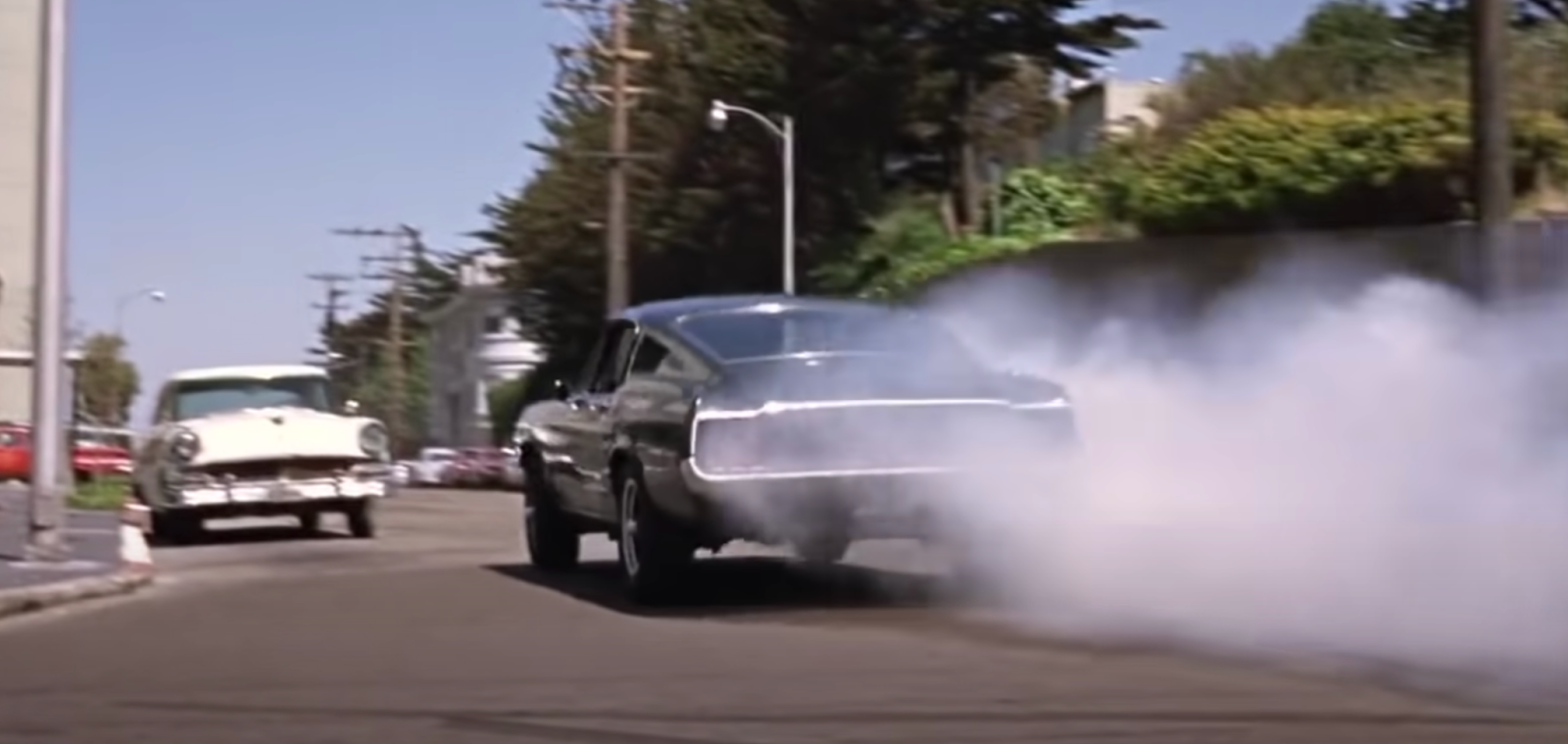}}
            \label{fig:}
    \end{subfigure}
    \hfill
    \begin{subfigure}{.2\textwidth}
    \centering
            \fbox{\includegraphics[height = 2cm]{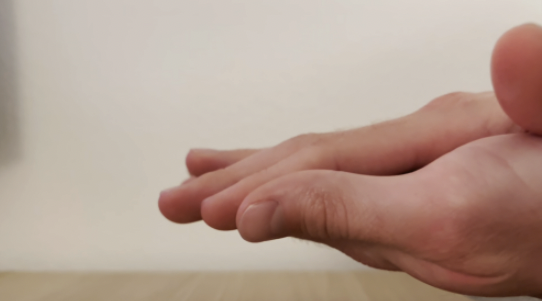}}
            \label{fig:}
    \end{subfigure}
    \hfill
    \begin{subfigure}{.2\textwidth}
    \centering
            \fbox{\includegraphics[height = 2cm]{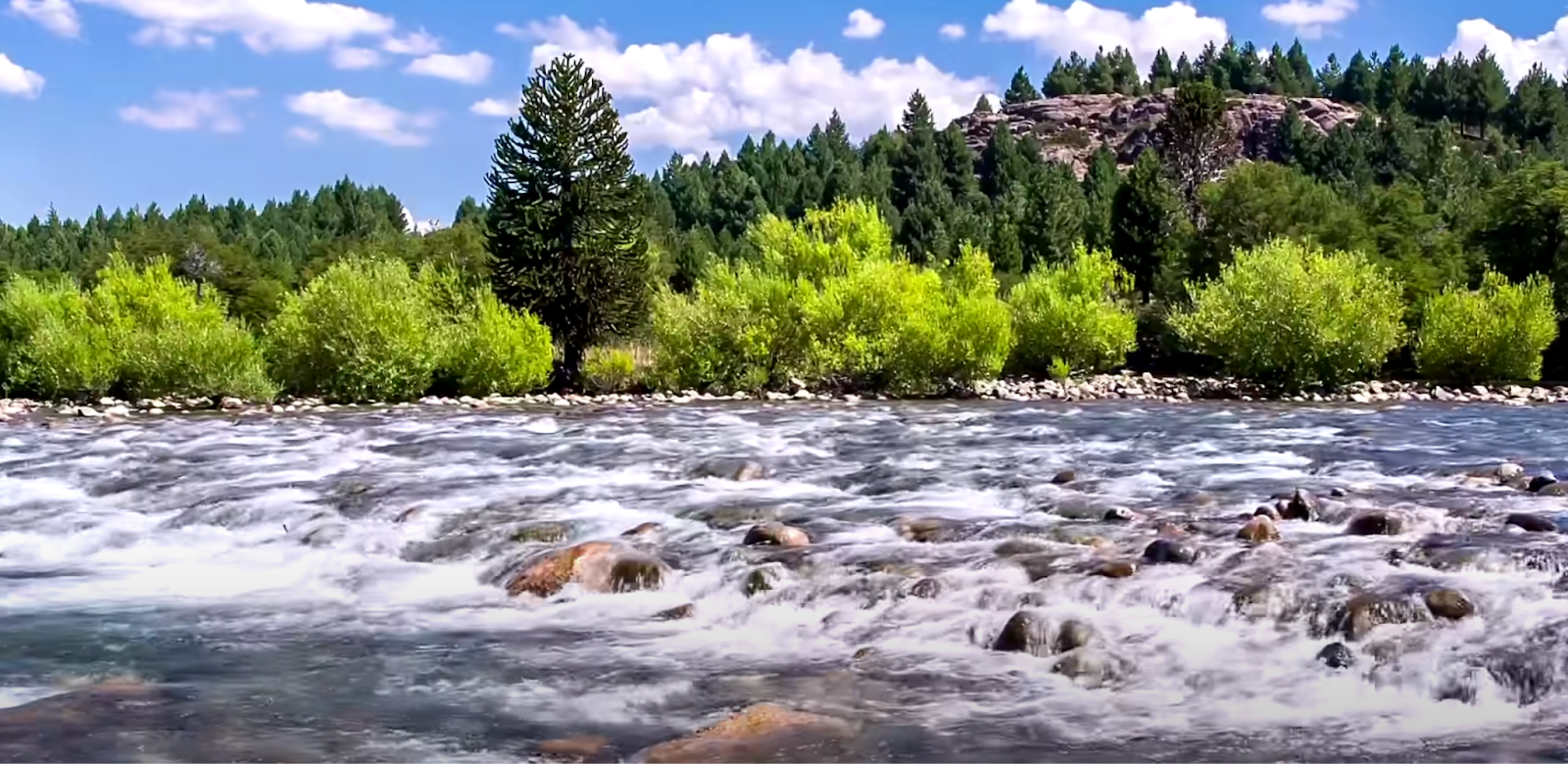}}
            \label{fig:}
    \end{subfigure}
    \hfill
    \begin{subfigure}{.2\textwidth}
    \centering
            \fbox{\includegraphics[height = 2cm]{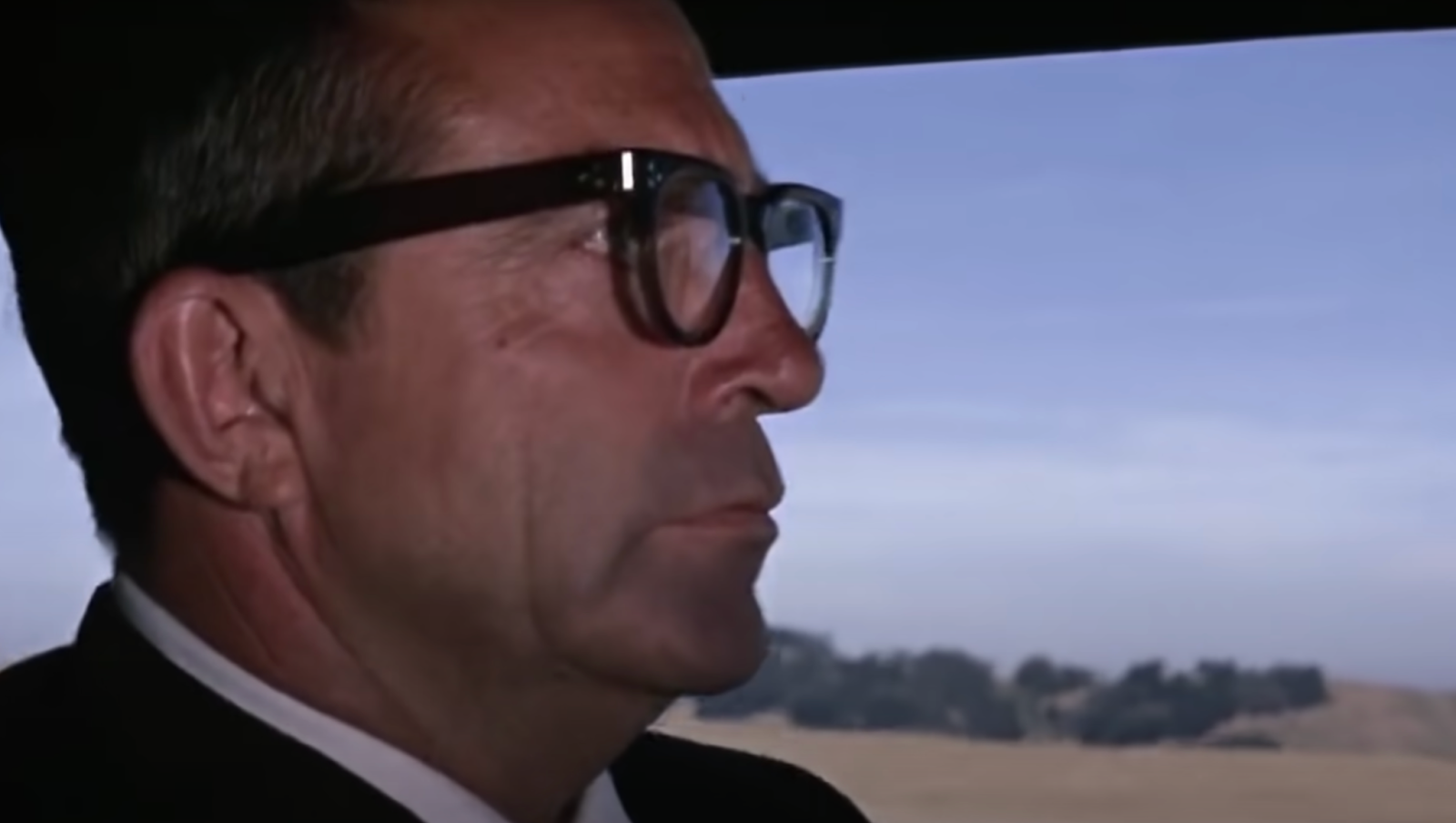}}
            \label{fig:}
    \end{subfigure}
    \caption{Example video frames from YouTube and self collected video data.}
\end{figure}

\begin{figure}[h!]
\centerline{\includegraphics[width=9cm]{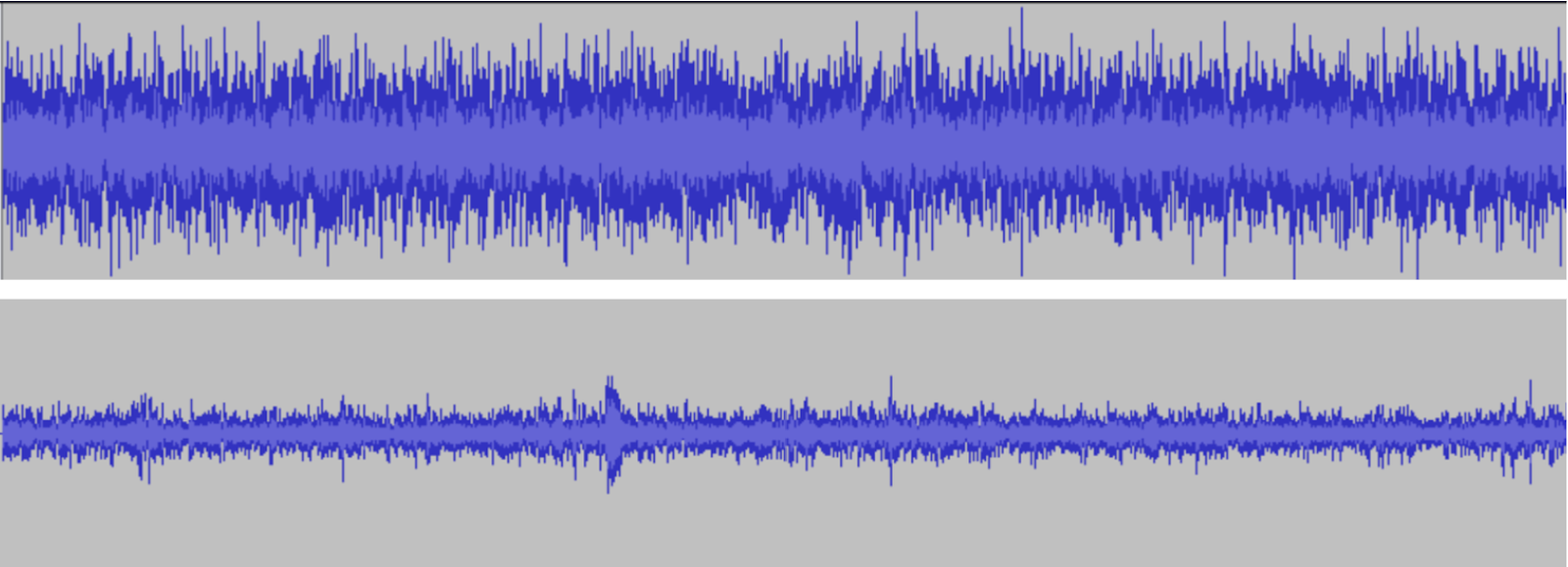}}
\caption{Example two channel audio data collected from a YouTube video.}
\label{fig:waveform}
\end{figure}

\subsection{Data Processing}
Processing and down sampling of the raw data was necessary due to memory and computational constraints. Audio data is read and down sampled using the Python library moviepy. Video data is read using the library scikit-video and individual frames are resized using the library cv2. Video and audio data are collected into a combined dataset which links frames from the video with corresponding audio sequences. When sampling out of this dataset any missing context is filled with zero padding to achieve a constant size (e.g., when generating audio for the first video frame). 

To account for discrepancies in the length of audio and video frame arrays (i.e., last video frame does not have expected number of associated audio samples given sample rate), we clipped the audio array by the length of the audio array modulo the expected audio samples per frame. We then clipped the video frame array such that its length was equal to that of the audio array divided by the expected audio samples per video frame. 


\subsection{Video Context Embedding} \label{vid_context_embedding}
\begin{figure}[h!]
\centerline{\includegraphics[scale=0.6]{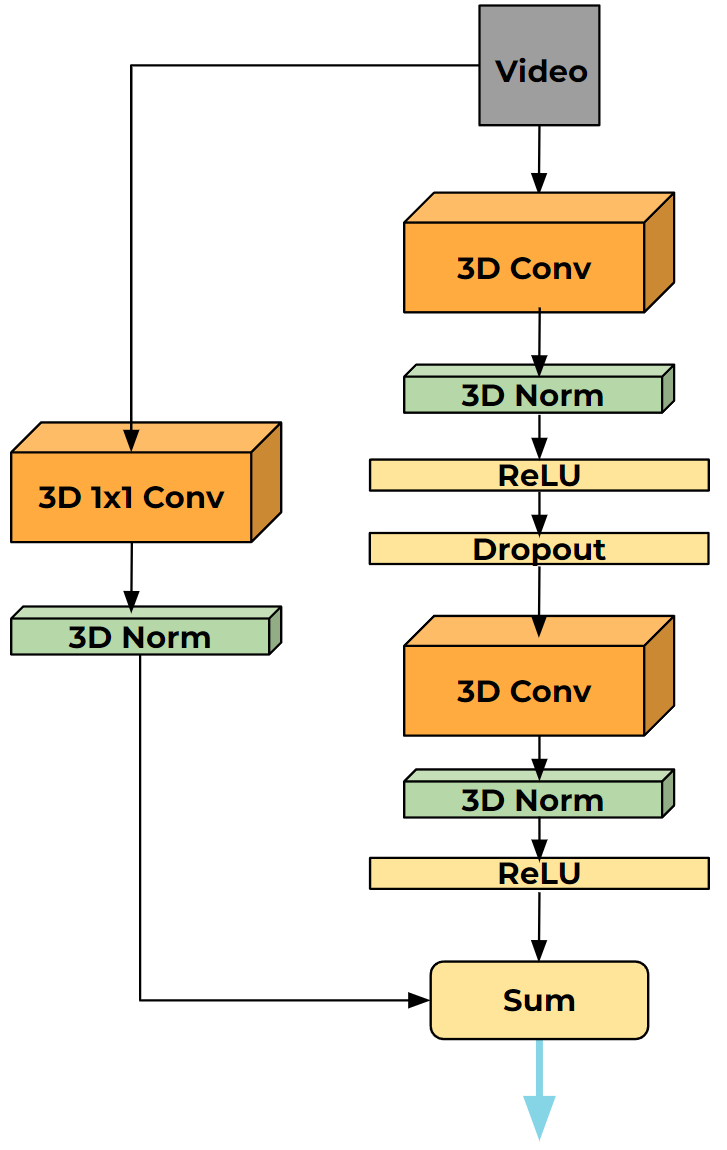}}
\caption{3D Residual Block Architecture}
\label{fig:resblock}
\end{figure}

To present our models with video context, we generated video context embeddings to describe the last $n$ frames using a ResNet 3D CNN architecture. The residual block architecture utilized for the video context embedding is illustrated in Figure \ref{fig:resblock}. A stack of these residual blocks is used to process the video input, before being projected into the dimensions of the audio data to form the video context embedding. The video to audio projection involves the flattening and projection of the video image dimensions to the audio sequence dimensions using a linear layer, $(x,y) \rightarrow (n_{aud})$, followed by projection of video channels to audio channels $(c_{vid}) \rightarrow (c_{aud})$ using a $1 \times 1$ convolution. This embedding is later fused with the audio data during model processing.

\subsection{Audio Generation}
\begin{figure}[h!]
\centerline{\includegraphics[scale=0.7]{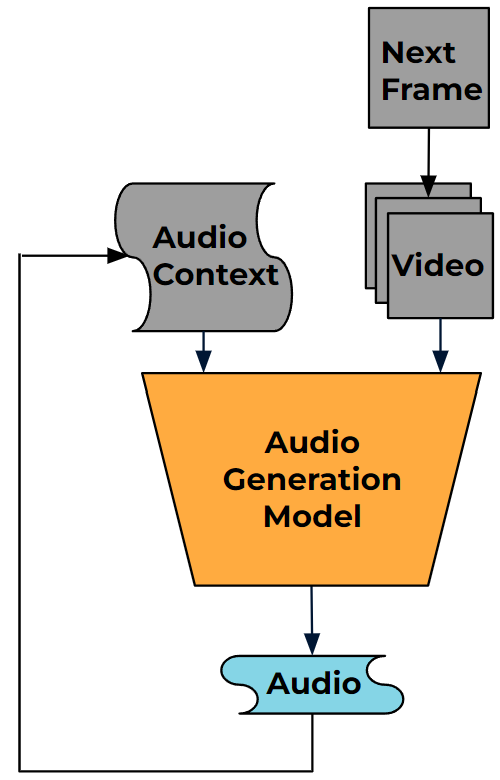}}
\caption{Audio Generation Process}
\label{fig:generation}
\end{figure}

During the audio generation process the model is fed with both an audio and video context. At each step the model is passed in all previously generated samples up to the context length, with any unfilled samples being zero padded. Similarly, video context is also zero padded if preceding video frames are unavailable. Audio output at each step is then stored until generation is complete, at which point the raw audio array is written to a MP3 output file. If the audio generation model being used generates audio element-wise rather than entire sequences the same video context must be used for multiple audio generation steps. The exact number of steps that the video context is frozen depends on the ratio of audio samples to video frames. As the audio being generated is two channel, the audio outputted at each step is either a pair or $(n,2)$ two dimensional sequence of floats between -1 and 1. This process is illustrated in Figure \ref{fig:generation}.\\

\subsubsection{Deep fusion architecture}
\begin{figure}[h!]
\centerline{\includegraphics[scale=0.4]{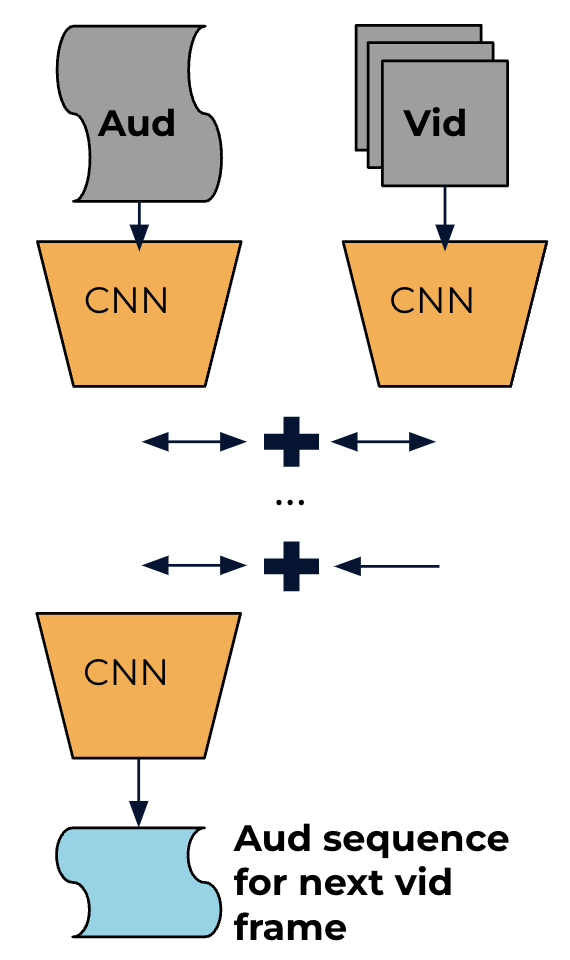}}
\caption{The deep fusion model for audio generation processes audio and video context in parallel and adds audio and video embeddings together with learned weights according to a data transformation procedure after each successive convolutional block. The output of a forward pass is the sequence of audio that corresponds to the next video frame. }
\label{fig:deep_fusion}
\end{figure}

The deep-fusion based architecture we develop is unique among the three architectures we test in that it outputs entire audio sequences associated with respective video frames rather than individual audio samples. The architecture takes in audio and video context for the past $n$ video frames, and processes them in parallel through ResNet CNN sub-blocks. Outputs from the audio and video processing streams are then transformed and added to each other after each sub-block. The technique used to transform video to audio was described in Section \ref{vid_context_embedding}. The audio to video transformation is similar to the video to audio transformation. This transformation begins with a projection from the audio dimension (sequence length) to video image dimension $(n_{aud}) \rightarrow (x,y)$ using a linear layer, followed by a channel projection to convert from the number of audio channels to the number of video channels using a $1\times1$ convolution $(c_{aud}) \rightarrow (c_{vid})$. The output of this layer is then reshaped into the video shape and can be merged with the video data. See Figure \ref{fig:deep_fusion} for an illustration of the end-to-end deep-fusion based architecture. 


\subsubsection{Dilated Wavenet CNN architecture}
\begin{figure}[thbp]
\centerline{\includegraphics[scale=0.4]{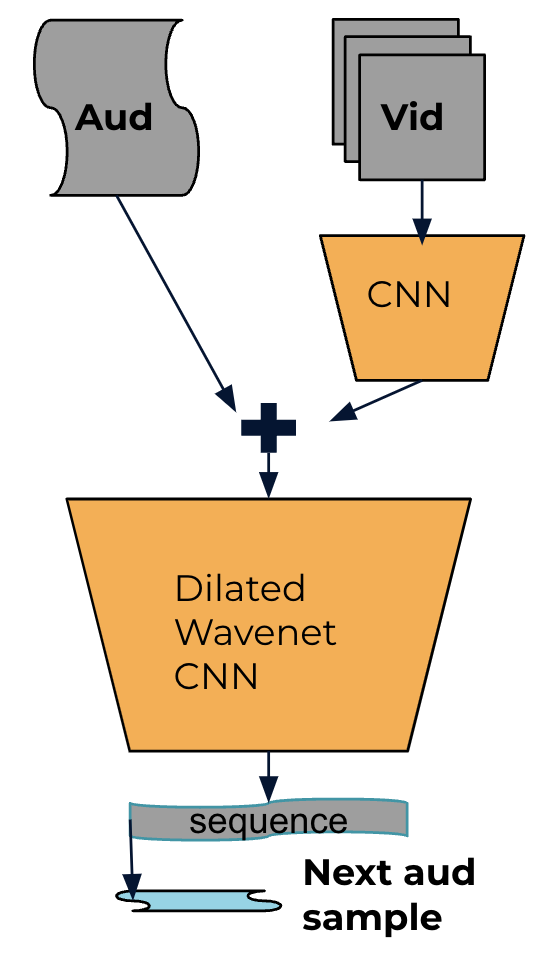}}
\caption{The Wavenet-based architecture applies the dilated CNN method from the original Wavenet paper \cite{wavenet} to an audio sequence added to a video embedding.}
\label{fig:wavenet_architecture}
\end{figure}

The dilated Wavenet CNN architecture is based on that described in the 2016 Wavenet paper \cite{wavenet}. The architecture consists of a casual dilated convolutional block followed by a stack of residual 1D convolutional blocks. Finally, outputs are passed through a dense layer followed by a Tanh activation. The model outputs an audio sequence for which the last element is taken as the next audio sample. The input to this architecture is the raw audio context added to the video context embedding, which are combined using the video to audio transformation method described in Section \ref{vid_context_embedding}. Our implementation modifies the Wavenet implementation found here \cite{wavenet_github} to take in two-channel audio, as opposed to an array with rows (time, amplitude, channel). We also modify the output to produce two channel audio using a dense layer followed by a Tanh activation. Find an illustration of the end-to-end architecture in Figure \ref{fig:wavenet_architecture}. 

\subsubsection{Transformer}
\begin{figure}[thbp]
\centerline{\includegraphics[scale=0.4]{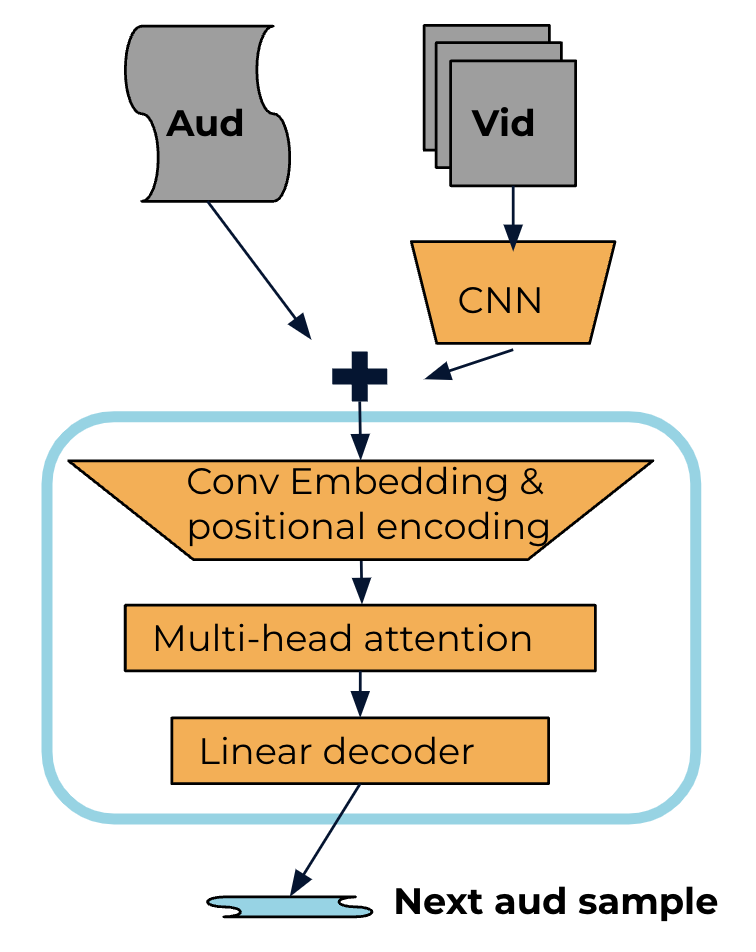}}
\caption{The Audio and Video transformer architecture takes in audio context added to a video context embedding, and uses a transformer module (outlined in blue) to output the next audio sample.}
\label{fig:transformer_arch}
\end{figure}

The third architecture we experimented with takes the same high-level structure as that of the Wavenet-based architecture, but differs in that it replaces the dilated Wavnet CNN block with a multi-head attention transformer module with implementation resembling that from the 2017 paper, "Attention is all you need" \cite{attention}. The module takes in a video embedding added to audio context using our video to audio transformation technique, which then receives a learned positional embedding. We tried two approaches to the audio context using the transformer model. The first approach utilized a series of strided convolutional layers to generate an audio embedding for the audio sequence before adding the video embedding. This served the purpose of further down sampling the audio to make training tractable for large context sizes. The second approach used smaller amounts of the raw audio data as the audio context with reduced transformer parameters. The model utilizes a linear decoder configured such that the next individual audio sample is output for each forward pass. ReLU activations are used throughout the model with a final Tanh activation at the end. Find an illustration of the end-to-end transformer-based architecture in Figure \ref{fig:transformer_arch}.


\subsubsection{Loss function} \label{loss_func}
All three models described above were trained with MSE, MAE, and cross-entropy loss, and were subsequently evaluated. Cross-entropy loss yielded the only reasonable results, and so it was used as the primary loss function for more refined model experimentation. Cross entropy loss is defined as follows:
$$H(P, Q) = - \sum_{x \in X} P(x) * log(Q(x))$$
where P represents the audio output distribution and Q represents the target distribution.

\section{EXPERIMENTS}
In addition to numerous qualitative experiments through which models were developed and refined, we evaluate final versions of our models using three different videos, each with unique qualities. The first video is a car chase scene with a lot of engine sounds, requiring models to capture mid-frequency patterns. The second is a home video of hands clapping for which success lies primarily in mirroring the low-frequency domain. The third video is one of various nature scenes with different types of sounds produced by various water features (e.g., rivers, breaking ocean waves, waterfalls). To mimic these noises a model must mimic a wide range of audio frequencies. 

We train each of our models on these three videos and subsequently have them generate audio for unseen samples of the same videos. We qualitatively evaluate the resulting audio outputs when paired with the video reserved for testing. We also compare our different methods by using validation loss as calculated via the cross entropy loss function described in Section \ref{loss_func} as a measure of "how close" the waveforms our models return are to those of ground-truth audio. Find a summary of the loss results in Table \ref{tab:loss_summary}. 

\begin{table}[h]
\centering
\begin{tabular}{l | l | l | l}
Test Video  & Deep Fusion & Wavenet-based & Aud \& Vid Transformer \\
\hline
Car chase & 1.65133e-05 & -0.03785 & -0.22000 \\
Clapping &  -1.36272e-07 & 0.00029 &  -0.00797\\
Nature & 1.04321e-05  & 0.01669 & -0.00862
\end{tabular}
\caption{Validation cross-entropy loss for the various models trained on different videos. Figures represent the best result recorded after hyper-parameter tuning. The Aud \& Vid transformer architecture demonstrates the lowest loss, and has the best performance in practice.}
\label{tab:loss_summary}
\end{table}

\subsubsection{Deep Fusion CNN Experiments} 
\begin{figure}[thbp]
\centerline{\includegraphics[scale=0.4]{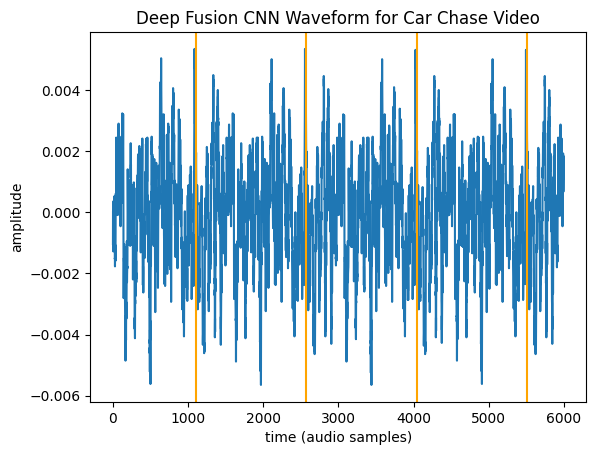}}
\caption{The waveform generated by the deep fusion CNN model suffered from discontinuity between audio segments generated for adjacent video frames. See a sample waveform generated by this model in blue, and markers indicating the beginning of new video frames in orange. Significant spikes in the waveform correspond with each new frame, adding a dominant, unwanted frequency.}
\label{fig:df_waveform}
\end{figure}

Experiments using our deep-fusion CNN model were largely unsuccessful. Models following this structure failed to generate sound that correlated with video features. Further, discontinuity between audio sequences associated with each video frame introduced a noticeable dominant frequency in many cases. See Figure \ref{fig:df_waveform} for an illustration. In other cases, the deep fusion CNN produced waveforms with values very close to zero. Examining the test cross-entropy loss results in Table \ref{tab:loss_summary}, is no surprise that the validation loss values associated with this model are higher than those associated with the Audio and Video Transformer model, for example, as this model was able to generate audio that captured some low and medium frequencies associated with input video. 

\subsubsection{Wavenet-based Architecture Experiments} 
\begin{figure}[thbp]
\centerline{\includegraphics[scale=0.4]{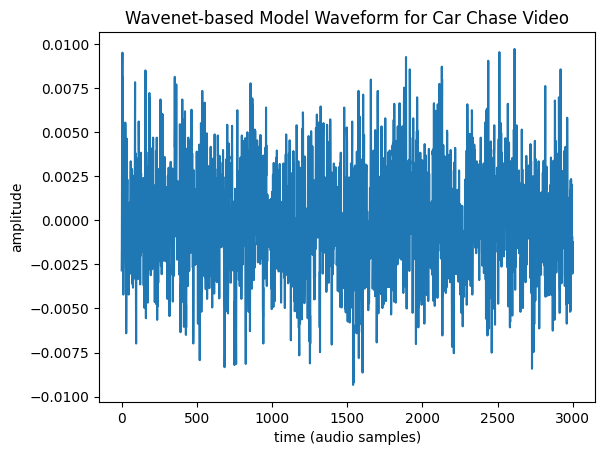}}
\caption{Audio waveform returned by Wavenet-based model for car chase video resembled white noise.}
\label{fig:wn_aud_sample}
\end{figure}

The Wavenet-based model returned complex audio that captured a wide frequency range. It achieved a better loss value than the deep fusion model for the car chase video, but not for the clapping or nature videos. Its outputs resembled white noise and did not demonstrate a clear correlation to video events. Outputs notably did not include frequencies correlated with frame rate, an improvement over the deep-fusion model outputs. This being said, the white-noise generated by the Wavenet-based model did resemble common tones heard throughout the car-chase video in particular, as compared by listening to generated audio and ground truth audio in succession. Find an example wavefrom generated by this model for the car chase video in Figure \ref{fig:wn_aud_sample}. Outputs for the clapping and nature scenes were consistently very close to zero.   

\subsubsection{Audio \& Video Transformer Experiments}
\begin{figure}[thbp]
\centerline{\includegraphics[scale=0.35]{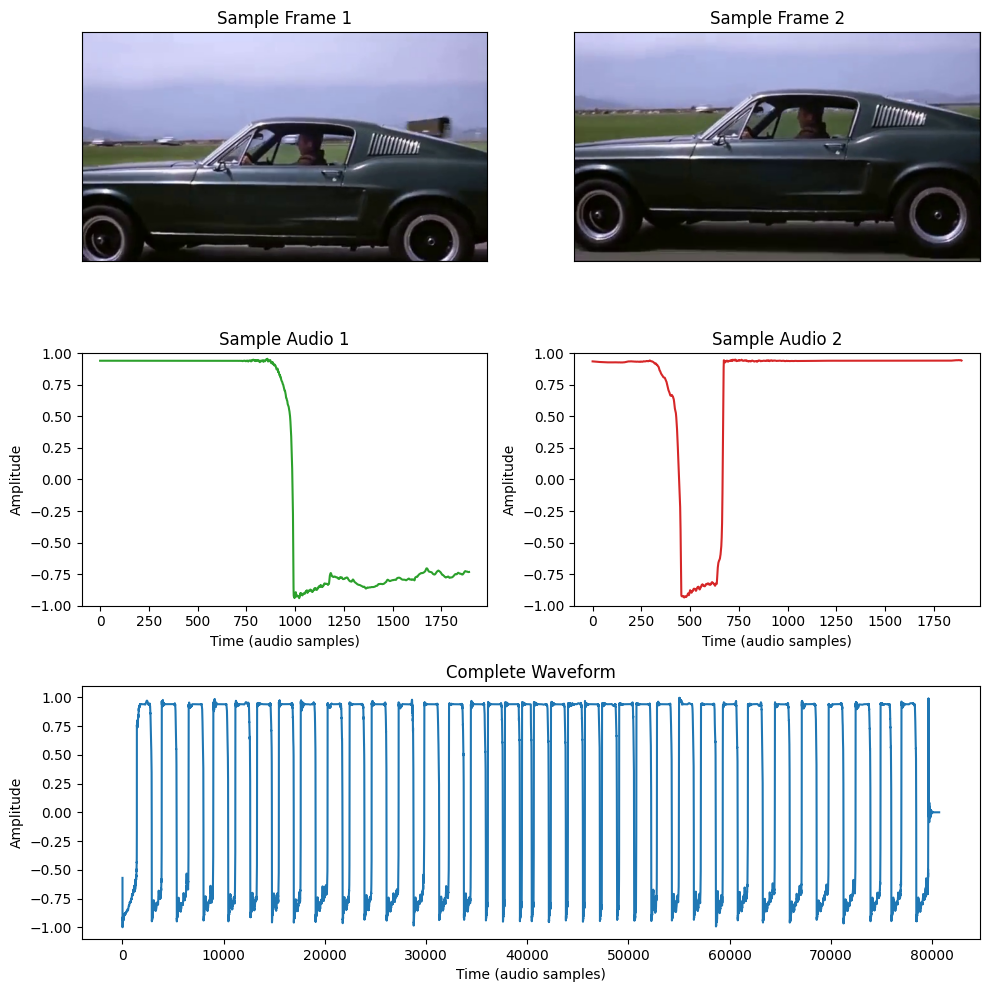}}
\caption{The waveform generated by the Audio \& Video Transformer for a clip from the car chase video. Two sample frames and their accompanying audio are displayed along with the entire waveform generated for the video clip.}
\label{fig:car_waveform}
\end{figure}

\begin{figure}[thbp]
\centerline{\includegraphics[scale=0.5]{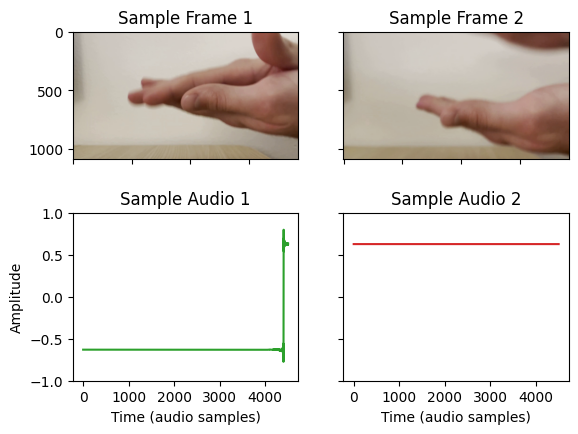}}
\caption{The Audio \& Video transformer correctly predicted audio discontinuities upon the incidence of hands during a clap, and flat audio in the absence thereof. However, it learned an odd resting position (+/- 0.6) which is perceptually equivalent to the resting position of 0.0 found in the original video, but nonetheless different.}
\label{fig:clap_waveform}
\end{figure}

Variations of our audio and video transformer architecture produced the best results when compared to the other techniques we developed. This model produced the most perceptually accurate results when paired with live video, and also yielded the lowest validation loss, suggesting the waveforms generated using this method were the closest to the true video waveforms. See Table \ref{tab:loss_summary} for a summary of loss results for this architecture. The loss associated with the car chase video was particularly low, and generated audio sounded quite similar to the muscle-car engine noise present in the original video. In Figure \ref{fig:car_waveform}, the regular waveform pattern associated with the engine noise is illustrated, along with a slight shift in waveform that takes place when the camera pans slightly, resembling a common effect seen throughout this video. In another example, this model was able to recognize instances of clapping and generate sound discontinuities to pair with them. See Figure \ref{fig:clap_waveform} for an illustration. Outputs for the nature video were very close to zero. We also attempted to modify this architecture by using quantized output between $[0,255]$ to encode audio amplitudes instead of using continuous output values. This transformer model then used a softmax activation to pick the most likely next audio element. Unfortunately, this model didn't end up performing well during validation and that line of investigation was dropped due to time constraints.

\section{CONCLUSION}

We presented three different model architectures for generating audio from silent video. Specifically, we developed a deep-fusion CNN architecture, Wavenet with added visual context architecture, and an audio and video transformer architecture. Our key idea was to generate audio iteratively by observing past audio and video context, in addition to the current video frame for each respective generation step. Past audio context consisted of audio generated by our models in previous steps. Of the three architectures we found the transformer-based method to be the most successful, as it was able to reasonably generate low and mid frequencies for some videos.

There are several key insights we can derive from this work. First, our work demonstrates that using state-of-the-art audio generation methods, like the recent transformer architecture for generating audio presented in \cite{transformer_sound}, is reasonable even when introducing video context. While these architectures have been primarily tested in more specific audio generation tasks (e.g., piano music synthesis), they may also generalize to a greater variety of sound generation tasks given new datasets. This insight can be seen in the results we demonstrate using our audio and video transformer architecture, which was able to generate low and mid frequency audio similar to that expected from a given video. Further, our work also illustrates the efficacy of the sample-by-sample approach to audio generation as opposed to sequence-by-sequence. Our audio and video transformer took a sample-by-sample approach, generating only a single audio sample at a time. In doing so, it was able to generate smooth audio with spikes only at key moments (e.g., when hands came together to clap). In contrast, our deep-fusion CNN generated audio sequences for each frame of a video. This lead to the addition of dominant, unwanted frequencies coinciding with video frame transitions. 

Second, the work we present reinforces the intuition that presenting an audio-from-audio generation model with additional video context in the beginning stages of the forward pass can lead to audio generation that mirrors video context. While not a direct comparison, we can see that the results from our transformer architecture, where video embeddings were introduced as an initial step, were superior to those of our deep-fusion cnn, where video context was introduced to varying degrees several times in the forward pass. This idea makes sense intuitively, because one would think it would be helpful to understand the scene you are predicting audio for before you begin to do the prediction. 

Finally, our work demonstrates that video context certainly can inform audio generation, but not without significant limitation. Our best models were able to predict dominant, low-to-mid frequency patterns present in the data (e.g., audio spikes associated with hands clapping, motor sound associated with cars), but failed to capture more-nuanced sounds and high-frequencies. One possible contributing factor is that these more nuanced sounds (e.g., the echo after a very loud clap) are not encoded by the video context in a meaningful way (to continue with the same example, the hands are not touching during the echo, so why would there be sound?). Another possibility is that in down sampling the training audio the high frequency information is lost, resulting in a model unable to replicate the original sound. In order to truly address this problem, our audio-generation modules will need to be able to extrapolate more from audio alone, which is perhaps possible to achieve with extensive training data and more model parameters. 

When considering the implications of our findings, it is important to note the limitations of our work. During this initial exploration phase, our models were trained on one video at a time to predict audio for unseen segments of the same video. This approach was used to accelerate the exploration cycle, as audio generation models can take exceedingly long to train (our models still needed to train over-night in many cases even with this approach). Thus, our successful models are over-fitted to specific types of video data, and likely do not generalize well to data outside of the limited distribution they have observed. This being said, we believe that our findings from this initial testing can point us in the right direction for future model development. 

Our vision for future work consists of two primary areas of exploration. First, we are optimistic that scaling up our audio and video transformer model and training it on a larger, more diverse dataset may yield interesting or promising results. Specifically, we are interested in testing an architecture more comparable to that developed in \cite{transformer_sound}, perhaps extended with additional parameters or additional transformer blocks to accommodate the additional complexity of our task. With a larger, more diverse dataset, and much longer training periods, we hope we will be able to successfully utilize the full capacity of a model like this. 

Second, we expect that additional experimentation with our video context embedding technique may yield positive results. One idea is to develop a model that can learn to attend to different frames, or different components of different frames in the video context when developing the video context embedding. This is opposed to our current method, where attention is applied after the video embedding has been added to the audio context.


\section{APPENDIX}
Find example code on \href{https://github.com/jaxwagner/sound_from_video}{github}

Find example demo videos at this \href{https://docs.google.com/presentation/d/1DQyNuZ0tQ7raM4nFebEV92pv9TvV66l2aJIzngrCGfA/edit#slide=id.g2405529ef5d_0_17}{link}

\bibliographystyle{bibliography/IEEEtran}

\end{document}